
\newskip\oneline \oneline=1em plus.3em minus.3em
\newskip\halfline \halfline=.5em plus .15em minus.15em
\newbox\sect
\newcount\eq
\newbox\lett
\newdimen\short
\def\adv{\global\advance\eq by1}
\def\set#1#2{\setbox#1=\hbox{#2}}
\def\nextlet#1{\global\advance\eq by-1\setbox
                \lett=\hbox{\rlap#1\phantom{a}}}

\newcount\eqncount
\newcount\sectcount
\eqncount=0
\sectcount=0
\def\sectadv{\global\advance\sectcount by1    }
\def\secta{\global\advance\sectcount by1    }
\def\equn{\global\advance\eqncount by1\eqno{(\copy\sect.\the\eqncount)} }
\def\put#1{\global\edef#1{(\the\sectcount.\the\eqncount)}           }

\magnification = 1200{\rm}

\voffset 0.2 truein
\vsize   8.7 truein
\hoffset 0.25 truein
\hsize  6.3 truein
\hfuzz 30 pt

\def\mbox#1#2{\vcenter{\hrule \hbox{\vrule height#2in
                \kern#1in \vrule} \hrule}}  
\def\sq{\,\raise.5pt\hbox{$\mbox{.09}{.09}$}\,}
\def\sqb{\,\raise.5pt\hbox{$\overline{\mbox{.09}{.09}}$}\,}

\def\ga{\gamma}
\def\sig{\sigma}

\def\bR{\overline R}
\def\R{R^2}
\def\Ric{R_{ab}R^{ab}}
\def\Rie{R_{abcd}R^{abcd}}
\def\sqR{\sq R}
\def\eps{\epsilon}

\def\trp{\rm{Tr'}}
\def\hD{\hat D}
\def\hH{\hat H}
\def\h1{\hat 1}
\def\hP{\hat P}
\def\hR{\hat{\cal R}}

\def\cPoly{1}
\def\cDuf{2}
\def\cWZ{3}
\def\cAntMot{4}
\def\cOdR{5}
\def\cMM{6}
\def\cAnt{7}
\def\cBV{8}
\def\cDeW{9}
\def\cBlau{10}
\def\cFV{11}
\def\cCD{12}
\def\cScmh{13}
\def\cBC{14}
\def\cFT{15}

\rightline{CPTH-A173.0492}
\rightline{LA-UR-92-1483}
\rightline{April 1992}
\vskip 1.5truecm
\centerline{\bf {CONFORMAL SYMMETRY AND CENTRAL CHARGES IN 4 DIMENSIONS }}
\vskip 2truecm
\centerline{Ignatios Antoniadis}
\centerline{Centre de Physique Theorique}
\centerline{Ecole Polytechnique}
\centerline{91128 Palaiseau, FRANCE}
\vskip .5truecm
\centerline{Pawel O. Mazur}
\centerline{Dept. of Physics and Astronomy}
\centerline{University of South Carolina}
\centerline{Columbia, SC 29208}
\vskip .3truecm
\centerline{and}
\vskip .3truecm
\centerline{Emil Mottola}
\centerline{Theoretical Division, T-8}
\centerline{Mail Stop B285}
\centerline{Los Alamos National Laboratory}
\centerline{Los Alamos, NM 87545}
\vskip 1truecm \centerline{\bf Abstract}
\vskip .5truecm
The trace anomaly of matter in curved space generates an
effective action for the conformal factor of the metric tensor
in $D=4$ dimensions, analogous to the Polyakov action for
$D=2$. We compute the contributions of the reparameterization
ghosts to the central charges for $D=4$, as well as the quantum
contribution of the conformal factor itself. The ghost contribution
satisfies the necessary Wess-Zumino consistency condition only
if combined with the spin-2 modes, whose contributions to the
trace anomaly we also discuss.

\vfill
\eject

\newcount\eqncount
\sectadv
\set\sect{1}
\beginsection{1. Introduction}

\par
In recent years there has been a great deal of interest in conformal
field theory (CFT) in two dimensions. Most of this attention has focused
on the strictly two dimensional aspects of CFT, such as the existence of an
infinite dimensional Virasoro algebra, characterized
by a single number, the central charge.
In critical string theory the matter central charges are cancelled by the
$-26$
contribution of the reparameterization ghosts, and the two dimensional
metric of
the world sheet decouples from the CFT matter system. Since the Einstein
action
in two dimensions is a topological invariant, the 2D metric has no (local)
dynamics in critical string theory and may be neglected.  \par In
non-critical
string theory, the  metric does not decouple and must itself treated as a
dynamical degree of freedom on an equal basis with the matter in the full
quantum theory.  In this case the relationship of the central charge of the
Virasoro algebra  to the trace anomaly of 2D gravity becomes apparent. In
the
covariant  approach to string theory initiated by Polyakov[\cPoly], the
trace
anomaly induces a non-local covariant effective action for 2D gravity which
becomes  local in the conformal parameterization (gauge),
$$
g_{ab}(x) = e^{2
\sig(x)} \bar g_{ab}(x). \equn\put\confdef
$$
Here $\bar g_{ab}(x)$ is a
fiducial metric on a surface of fixed topology. The Polyakov-Liouville
theory
for $\sig(x)$ describes the dynamics of fluctuating random surfaces in two
dimensions. In four dimensions,  the quantum dynamics of the conformal
factor
may be studied in an analogous manner by constructing the effective action
generated by the trace anomaly[\cAntMot]. Conformal field theory techniques
then
provide information about the long distance behavior of four dimensional
quantum
gravity directly. \par   Proceeding by analogy with two dimensions, we
begin by
considering a CFT with matter fields, denoted generically by $\phi_i$,
which
transform with definite conformal weights $\alpha_i$, and described by the
classical conformally invariant action,
$$
S_{cl}[e^{-\alpha_i \sig} \phi_i;e^{2
\sig} \bar g_{ab}] = S_{cl} [\phi_i;\bar g_{ab}].
$$
At the quantum level this symmetry of the matter action is broken by the
trace anomaly. The general form of the quantum trace anomaly of a CFT in
curved
space is:
$$
\eqalign{
\langle T_a^{\ a}\rangle &= {2 \over \sqrt{-g}}g_{ab}{\delta S_{eff}
\over \delta g_{ab}} [\phi_i = 0; g_{ab} = e^{2 \sig} \bar g_{ab}] \cr
& = b F + b' (G - {2 \over 3} \sq R) + {\zeta \over 3} \sq R \cr & \equiv
{1
\over \sqrt{-g}}{\delta \over  \delta \sig}\Gamma[\bar g_{ab}; \sig] ,\cr}
\equn\put\anomf
$$
where the notation,
$$
F \equiv C_{abcd}C^{abcd} = \Rie - 2 \Ric + {1 \over 3} \R \equn\put\weyl
$$
for the square of the Weyl tensor and
$$
G \equiv \Rie - 4 \Ric + \R \equn\put\gaussb
$$
for the Gauss-Bonnet integrand is employed. If non-zero background matter
fields are considered, additional terms proportional to
matter field operators, multiplied by the beta functions of the
corresponding dimensionless couplings will appear in {\anomf}.
\par
The coefficients $b$ and $b'$ have been calculated for free CFT's
in four dimensions[\cDuf]:
$$
\eqalign{
b &=\ \ {1 \over (4 \pi)^2}{1 \over 120}(N_S + 6\ N_F + 12 N_V -\ \ 8) +
b_G,\cr
b'&= -{1 \over (4 \pi)^2}{1 \over 360}(N_S + 11 N_F + 62 N_V - 28) +
b'_G,\cr}
\equn\put\bcoef
$$
where $N_S$, $N_F$, and $N_V$ are the numbers of conformally coupled
scalar,
Dirac spinor, and vector gauge fields respectively. The additional
contributions
to these coefficients given in {\bcoef} are the quantum $\sig$ and
gravitation/ghost contributions respectively which we discuss below. They
are
the direct analogues of the $+1$ and $-26$ contributions to the central
charge
in the two dimensional Polyakov theory. Unlike $b$ and $b'$, the
coefficient of
the $\sq R$ term of the anomaly is altered by the addition of a {\it local}
$R^2$ term in the action, so that it must be treated as an additional
renormalized coupling, and the $\zeta$ coefficient is left undetermined.
\par
An $R^2$ term in the anomaly is forbidden for CFT's by the Wess-Zumino (WZ)
consistency condition[\cWZ]. In the present context this is simply the
statement
that the variational relation {\anomf} is integrable,  {\it i. e.} that
there
exists an effective action functional $\Gamma$, whose $\sig$ variation is
the
anomaly, such that the full effective action depends only on the complete
metric
in {\confdef}:
$$
S_{eff}[g_{ab}= e^{2 \sig} \bar g_{ab}] = S_{eff}[ \bar
g_{ab}]  + \Gamma [\bar g_{ab}; \sig]. \equn\put\effact $$ By subjecting
the
background metric and conformal factor to the simultaneous transformation,
$$\eqalign{ \bar g_{ab} &\rightarrow e^{2 \omega}\bar g_{ab}\cr \sig
&\rightarrow \sig - \omega, \cr} \equn\put\weyl
$$
which leaves the total metric unchanged, we immediately deduce from
{\effact}
that the functional $\Gamma$ must satisfy the relation,
$$
\Gamma[{\bar g};\sig]=\Gamma[{e^{2\omega}\bar g};\sig-\omega]+
\Gamma[{\bar g};\omega], \equn\put\effg
$$
which is one form of the WZ condition.
By taking two successive conformal variations of {\effact} in
different orders, we arrive at a second form of the WZ condition:
$$
\Gamma[\bar g_{ab}; \sig_1] - \Gamma[e^{2 \sig_2}\bar g_{ab}; \sig_1] =
\Gamma[\bar g_{ab}; \sig_2] - \Gamma[e^{2 \sig_1}\bar g_{ab}; \sig_2]\equn
$$
{}From this relation, expanded to first order in $\sig_1$ and $\sig_2$, it
follows that the local variation of the trace anomaly must yield a
self-adjoint operator, symmetric in $\sig_1$ and $\sig_2$. One
can easily check that the variation of $R^2$ yields a non-self adjoint
operator (since $R \sq \neq \sq R$), so that an $\R$ term cannot appear in
the conformal anomaly {\anomf}, if {\effact} is satisfied.
\par
The existence of the relation {\effact} has another important consequence.
Consider the effective $\sigma$ theory defined by $\Gamma[{\bar g};\sig]$
and the conformally invariant measure $[{\cal D} \sig]_g = [{\cal
D}\sig']_g $,
where ${\sig}'=\sig-\omega$. We shall show now that the local Weyl
invariance
under {\weyl} which leaves the metric $g$ invariant
and the Wess-Zumino consistency condition imply that the {\it total}
conformal
anomaly of matter plus ghosts plus $\sigma$ {\it vanishes identically}.
\par
Proceeding in analogy with the two dimensional case, we begin by
considering
the form of the generally covariant functional measure on the (cotangent)
space
of infinitesimal metric deformations, $\delta g_{ab}$. This is defined by
the
Gaussian normalization integral,
$$
\int [{\cal D} \delta g]_g \exp \bigl\{ - {i \over 2} <\delta g, \delta
g>_g
\bigr\} = 1   \equn\put\norm
$$
The quadratic inner product on the space of metric fluctuations is defined
by:
$$
<\delta g, \delta g>_g \equiv \int d^4 x\ \sqrt{-g}\ \delta g_{ab}\
G^{abcd}
\ \delta g_{cd}   \equn\put\innr
$$
which makes use of the DeWitt supermetric[\cDeW],
$$
G^{abcd} = {1\over 2} ( g^{ac} g^{bd} + g^{ad} g^{bc} + C g^{ab} g^{cd} )
\equn\put\dew
$$
at the field ``point" $g_{ab}$. The form of $G^{abcd}$ is determined
up to the constant $C$  by the requirement that $G^{abcd}$ be ultralocal,
{\it i. e.} free of derivatives of $g_{ab}$. This definition of the
functional
measure on the cotangent space of metrics is generally covariant, and
depends only the full metric $g$. Hence it is unchanged by the Weyl
transformation {\weyl}. It induces a covariant definition of the functional
measure on the scalar conformal subspace of metrics parameterized by $\sig$
in
{\confdef}. Let us define the Jacobian ${\cal J}$ relating this $\sigma$
dependent covariant measure $[{\cal D} \sig]_g$ induced by {\norm} and
{\innr}
above to the translationally invariant, $\sig$ independent measure at the
point
$\bar g$ by the relation,
$$
[{\cal D} \sig]_g ={\cal J}[{\bar g ;\sig}][{\cal
D} \sig]_{\bar g}. \equn\put\conjac
$$
The implication of the fact that $[{\cal
D} \sig]_g$ depends only on the full metric $g$ is that this Jacobian must
satisfy:
$$
{\cal J}[e^{2\omega}{\bar g};\sigma-\omega]{\cal J}[{\bar
g};\omega] = {\cal J}[{\bar g};\sig].\equn
$$
{}From this it is clear that the
$\sigma$ theory can be defined with respect to the translationally
invariant
measure $[{\cal D} \sig]_{\bar g}$ with a new action, ${\Gamma}_1=\Gamma -
i \ln
{\cal J}$ which also satisfies the Wess-Zumino condition {\effg}. Therefore
${\Gamma}_1$ must have the same form as $\Gamma$, but with different
coefficients $b$ and $b'$ in {\anomf}. The action ${\Gamma}_1$ will be
determined completely once we prove that its conformal anomaly compensates
the
conformal anomaly of matter fields (and ghosts). The simplest way to prove
that
is the following. Consider the result of integrating out the $\sigma$
field:
$$
\eqalign{ e^{iW[{\bar g}]}& =\int [{\cal D} \sig]_g \ e^{i S_{eff}[g_{ab}=
e^{2 \sig} \bar g_{ab}]}\cr & = e^{i S_{eff}[ \bar g_{ab}]} \int [{\cal D}
\sig]_{\bar g e^{2 \sig}}\  e^{i\Gamma [\bar g_{ab}; \sig]}\cr & = e^{i
S_{eff}[
\bar g_{ab}]} \int [{\cal D} \sig]_{\bar g}\  e^{i\Gamma_1 [\bar g_{ab};
\sig]}.  \cr} \equn\put\fullint
$$
Now subject this functional to the
transformation {\weyl}, using the WZ condition {\effg} and translational
invariance of the measure $[{\cal D} \sig]_{\bar g}$:
$$
\eqalign{
e^{iW[e^{2\omega}{\bar g}]}& =e^{i S_{eff}[ e^{2\omega} \bar
g_{ab}]}\int[{\cal
D} \sig']_{\bar g}\  e^{i\Gamma_1[e^{2\omega}{\bar g}; \sig']} \cr & = e^{i
S_{eff}[ e^{2\omega} \bar g_{ab}]}e^{-i\Gamma[{\bar g};\omega]} \int [{\cal
D}
\sig]_{\bar g e^{2 \sig}}\  e^{i\Gamma[{\bar g};\sig]}\cr & =e^{i S_{eff}[
\bar
g_{ab}]} \int [{\cal D} \sig]_{\bar g}\  e^{i\Gamma_1 [\bar g_{ab};
\sig]},\cr}\equn
$$
where the substitution, $\sig' = \sig -\omega$ and {\effg}
has been used. Therefore,
$$
W[e^{2\omega}{\bar g}]=W[{\bar g}],\equn\put\anomcan
$$
and the full effective action after integration over
$\sig$ is Weyl invariant. \par Thus, the WZ condition implies that the {\it
total} trace anomaly must vanish. In other words, as a consequence of the
absence of diffeomorphism anomalies (which is the real physical content of
{\effact}), a form of local conformal invariance survives in the quantum
theory.
\par To proceed further, we notice that the $\Gamma$ satisfying the WZ
condition
may be found explicitly, by directly integrating {\anomf}[\cOdR]:
$$
\eqalign{
&\Gamma[\bar g;\sig] =  \ 2 b' \int d^4 x\ \sqrt{-{\bar g}}\ \sig
\overline\Delta_4 \sig + \int d^4 x\ \sqrt{-{\bar g}}\  \Bigl[ \ b\
\overline F
+ b'\ \bigl(\overline G - {2 \over 3} \sqb \bR \bigr)\Bigr] \sig \cr  &
\qquad
-{\zeta \over 36}  \int d^4 x \sqrt{-{\bar g}}R^2\vert_{g =e^{2\sig}
g_{ab}},\cr}\equn\put\actan
$$
where $\Delta_4$ is the Weyl covariant fourth
order operator acting on scalars:
$$
\Delta_4 = \sq ^2 + 2R^{ab} \nabla_a
\nabla_b - {2 \over 3} R {\sq}  + {1 \over 3} (\nabla^a R) \nabla_a .
\equn\put\fourth
$$
As in two dimensions, the $\sig$ dependence of the non-local
anomaly-induced action becomes local in the conformal parameterization
{\confdef}. However, it corresponds to a fully covariant but {\it
non-local} action,
$$
-{1 \over 8 b'} \int \bigl[ b F + b' (G - {2 \over 3} \sqR) \bigr]
{1 \over \Delta_4} \bigl[ b F + b' (G - {2 \over 3} \sqR) \bigr] -
{\zeta \over 36} \int R^2.\equn\put\nonf
$$
\par
This action plays a role similar to the WZ action of low energy pion
physics, as realized in the Skyrme model, for example. That is, it can
be interpreted as an effective action at low energies, which describes all
modifications to the Ward-identites due to the presence of the
quantum trace anomalies. The Ward identity corresponding to the local
Weyl transformation {\weyl} in the effective theory guarantees that
the total trace anomaly vanishes at the CFT fixed point. Contrary to the
higher derivative Skyrme model, however, when the $\sig$ action is treated
as a quantum action, we find that it is ultraviolet renormalizable.
Hence its renormalization group beta functions may be studied in ordinary
flat
space perturbation theory. The $\zeta$ coupling is infinitely renormalized,
and
therefore, would in general be expected to contribute an $R^2$ term to the
anomaly, proportional to $\beta(\zeta)$. Since WZ consistency requires that
there be no such term, general coordinate invariance of the effective
theory
requires that this $\beta$ function vanish, {\it i. e.} that we sit at a
fixed
point of the $\zeta$ coupling. In [\cAntMot] we found in flat space
perturbation
theory that $\zeta$ has an infrared stable perturbative fixed point at
$$
\zeta = 0. \equn\put\zefix
$$
In [\cAntMot] the fiducial metric was taken to be flat, and the residual
conformal Killing symmetries of flat spacetime appeared to play a
crucial role in the Ward identities and fixed point condition {\zefix}.
It now becomes evident that the fixed point conditions
on $\zeta$ and the other couplings of the $\sig$ action actually
derives from the WZ condition of eq. {\effact}, {\it i. e.} from the
requirement that the full effective action depend only on the combination
$g_{ab} = e^{2\sig}\bar g_{ab}$, for an {\it arbitrary} background
satisfying the full field equations of the theory.
The condition {\zefix} fixes the arbitrariness present in
$\zeta$, the coefficient of the local $R^2$ term in the effective action.
\par
The effective action {\actan} involves four derivatives of $\sig$ and
raises the problem of unitarity, known to plague local higher derivative
theories. However, recall that the Einstein action in its covariant form
appears
to contain a negative metric scalar degree of freedom, but that this
``degree of
freedom" is removed by the reparameterization ghosts in a covariant
framework[\cMM]. In the quartic action {\actan} there will remain one
additional
scalar dilaton degree of freedom not cancelled by the constraints. In order
to
settle the unitarity issue a study of the coordinate reparameterization
constraints, derived from the energy momentum tensor of the $\sig$ theory
must
be  undertaken, to determine if the negative norm states drop out of the
physical spectrum, as in $c > 25$ non-critical string theories[\cAnt]. The
results of this investigation will be reported in a separate publication.
\par
Once the action {\nonf} or {\actan} is added to the Einstein-Hilbert
action,
it describes dynamics of the conformal part of the metric tensor
which is quite different from the classical Einstein theory,
where $\sig$ is completely constrained by the classical equations of
motion.
In particular, the conformal field $e^{\sig}$, which has classical
scaling dimension one, receives an anomalous scale dimension $\alpha$
from the $\sig$ loops, given by the quadratic relation,
$$
\alpha = { 1 - \sqrt {1 - {4 \over Q^2}} \over {2 \over
 Q^2}}.\equn\put\alp
$$
where
$$
Q^2 \equiv 16 \pi^2 (\zeta - 2 b') = - 32 \pi^2 b' , \equn\put\centch
$$
by {\zefix}. The anomalous scaling of the conformal factor leads to
anomalous scaling of the Ricci scalar under conformal transformations as
well.
This observation we argued in [\cAntMot] is relevant to the issue of
the effective cosmological ``constant" of the low energy limit of
quantum gravity. The fact that the $\zeta = 0$ fixed point is infrared
stable
is essential for any application of the effective $\sig$ theory to long
distance physics in four dimensions. The contributions to $Q^2$ from
the $\Delta_4$ anomaly and reparameterization ghosts are required for any
such application as well. From eqs. {\bcoef} and {\centch} the
possibility that $Q^2$ could be less than or equal to four
remains open. From the anomalous scaling relation, {\alp} we observe
that the value $Q^2 = 4$ is a critical value
(corresponding to $c=1$ in two dimensions) at which
the theory could exhibit a phase transition with
qualitatively new phenomena. Thus, it would be very interesting if the
quantum $\sig$, ghost and gravitational contributions to this central
charge
could be calculated reliably, for eventual application to observations
of large scale structure in the universe. In this letter we describe
how these missing contributions quoted in {\bcoef} above may be obtained.

\vfill
\eject

\newcount\eqncount
\sectadv
\set\sect{2}
\beginsection{2. The Quantum $\sig$ Contribution to the Central Charges}

\par
We consider first the contribution of the quartic operator
$\Delta_4$ itself to the coefficients $b$ and $b'$. This is equivalent to
finding the modification of the effective action $\Gamma \rightarrow
\Gamma_1$
due to the shift from the coordinate invariant measure on $\sig$ induced by
the definitions {\norm} and {\innr} to the translationally invariant
measure
$[{\cal D} \sig]_{\bar g}$, in the language of the introduction. Notice
that
it makes no sense to define the Jacobian of this transformation in
{\conjac},
without reference to the operator $\Delta_4$, since it is the trace anomaly
of this operator which determines the difference $\Gamma_1 - \Gamma$ and
the full effective action $W[\bar g]$ in {\fullint}. This operator in four
dimensions is the precise analog of the scalar Laplacian in two dimensions
which
contributes to the central charge like one additional scalar degree of
freedom,
shifting $c_m - 26 \rightarrow c_m - 25$.  \par The classical Einstein and
cosmological terms are soft by comparison to the quartic term in $\Gamma$,
so
that they cannot affect the contribution of the $\sig$ field to $b$ and
$b'$.
Hence, we  consider the free $\sig$ action which is conformally invariant when
treated as a classical action since $\Delta_4$ transforms covariantly under
{\confdef}:
$$
\Delta_4 = e^{-4 \sig} \bar \Delta_4 . \equn\put\dvar
$$
The standard Seeley-deWitt heat kernel expansion does not apply
to this fourth order operator. However, a variety of other
methods have appeared in the literature to treat fourth order
operators. The general algorithm for the $a_2$ coefficient in
the heat kernel expansion of a minimal fourth order operator of the form,
$$
{\hat F} = \sq^2 \h1 + \hD^{(\mu\nu)}\nabla_{\mu} \nabla_{\nu} + \hH^{\mu}
\nabla_{\mu} + \hP \equn\put\oper
$$
is[\cBV]:
$$
\eqalign{
a_2 ({\hat F}) &= {1 \over (4 \pi)^2} {\rm tr} \Bigl\{{1 \over 90}
(\Rie - \Ric) \h1 + {1\over 6} \hR_{ab} \hR^{ab} \cr
&\qquad + {1 \over 36} \R \h1 - \hP - {1 \over 6} \hD^{ab} R_{ab}+ {1\over
12}
\hD R + {1 \over 48}\hD^2 + {1 \over 24} \hD_{ab}\hD^{ab}\Bigr\}\cr}.
\equn\put\algo
$$
In this algorithm $\h1$ denotes the unit matrix for the field upon which
the operator ${\hat F}$ acts and $\hR$ is the commutator of covariant
derivatives in the given representation:
$$\eqalign{
\h1^A_{\ B} \phi^B &\equiv \phi^A \cr
(\nabla_a \nabla_b - \nabla_b \nabla_a) \phi^A &\equiv \hR^A_{\ Bab}
\phi^B \cr\hD &\equiv \hD^{\mu}_{\mu}.\cr}
\equn
$$
Terms proportional to $\sqR$ have been neglected here in {\algo}.
\par
The scalar operator $\Delta_4$ appearing in {\fourth} is of this type with
$\hD^{ab} = 2 R^{ab} - {2 \over 3}Rg^{ab}$, $\hH^a = {1 \over 3}(\nabla^a
R)$,
and $\hP = \hR = 0$. Applying the general formula {\algo} to this case
yields the anomaly coefficient in the form of {\anomf} with the values
of $b$ and $b'$ equal to $-8$ and $-28$ respectively in scalar units,
as quoted in {\bcoef} above. In the basis of $F$ and $G$ used in {\anomf}
the coefficient of $R^2$ vanishes as required by the WZ condition for
the conformally covariant operator {\fourth} obeying {\dvar}. By using
this WZ condition we may check the result for the $b'$ coefficient
independently by the zeta function method, as follows.
\par
Consider the conformally flat and maximally symmetric $S_4$ with
radius $H^{-1}$ and vanishing Weyl tensor, $F=0$.
The spectrum of $\Delta_4$ on this space is $n(n+1)(n+2)(n+3) H^4$ with
the scalar degeneracy of ${1 \over 3} (n + 1) (n +
{3 \over 2})(n + 2)$. This yields the zeta function,
$$
\zeta_4 (s) = {1 \over 3} \Bigl( {\mu \over H } \Bigr)^{4s}
\sum_{n=1}^{\infty}{(n + 1)(n + {3 \over 2})(n + 2) \over
n^s (n+1)^s (n+2)^s (n+3)^s} . \equn\put\zedel
$$
The value of $\zeta_4 (0)$ determines the scaling behavior of
the operator $\Delta_4$. It may be determined by employing a
binomial expansion for each of the three factors in the denominator
of the form $(n + p)^s$, interchanging the sums generated by this
expansion with the $n$ sum, and performing the $n$ sum in terms of
the Riemann zeta function $\zeta_R$, whose analytic properties at
$s=0$ are well known. The result of this calculation is:
$$
\zeta_4 (0) = {1 \over 3} \zeta_R (-3) + {3 \over 2} \zeta_R (-2)
 + {13 \over 6} \zeta_R (-1) +  \zeta_R (0) - {1 \over 6} =
-{38 \over 45}. \equn
$$
To this we must add the one from the excluded mode at $n=0$, and
multiply the result by two to account for the fact that the
fourth order operator has twice the scaling behavior under $\mu
{d \over d \mu}$ as a second order operator to obtain the
result ${28 \over 90}$, or $-28$ in scalar units, as stated
in {\bcoef}. The same result is obtained more rapidly by considering
a general Einstein space,
$$
R_a^{\ b} = \Lambda \delta_a^{\ b}, \equn\put\eins
$$
which otherwise has no particular symmetry. Then the Riemann tensor
squared survives as an independent contribution to the anomaly
coefficient $a_2$. The quartic operator $\Delta_4$ factorizes
into the product of two second order operators in this case:
$\sq (\sq - {R \over 6})$. Since
the heat kernels of these scalar operators are well known,
the $b$ and $b'$ coefficients of $\Delta_4$ may be
determined by simply adding the results for these two
second order operators. Again the same results are
obtained.
\par
This factorization is instructive for a
different reason. It shows that the quartic action
may be regarded as containing a conformally coupled
scalar mode from $\sq - {R \over 6}$ which has a
negative a kinetic term, and a minimally coupled
scalar from $\sq$ with a positive kinetic term.
Nevertheless, the first operator contributes to both
$b$ and $b'$ like one ordinary conformally coupled
scalar, while it is the second operator which gives
the negative contributions ($-9$ and $-29$ in
scalar units) that makes the full contribution of
$\Delta_4$ negative. The first mode is similar
to the Liouville kinetic term in two dimensions with
$c>25$,  or the scalar mode of Einstein gravity whose
propagator must be cancelled by the ghosts if
the theory is to be unitary. The minimally
coupled scalar mode with positive kinetic term
has no analog in two dimensions or the Einstein
theory, and is responsible for the unusual behavior of this
theory in the infrared, discussed in [\cAntMot].

\vfill
\eject

\newcount\eqncount
\sectadv
\set\sect{3}
\beginsection{3. The Ghost Contribution to the Central Charges}

\par
Having determined the quantum contribution of the $\sig$
field itself to the anomaly coefficients, we turn now to
the more difficult problem of the ghost
contributions to the central charge(s). If in the definition of the
inner product {\innr} and covariant measure {\norm} we
decompose the general metric deformation in the form,
$$
\delta g_{ab} = h^{\perp}_{ab} + (L \xi )_{ab} + (2\sig + {1 \over 2}
\nabla
 \cdot \xi ) g_{ab}    \equn\put\defo
$$
where $L$ is the conformal Killing operator, mapping vectors into
traceless, symmetric tensors, and defined by:
$$
(L \xi)_{ab} = \nabla_a \xi_b + \nabla_b \xi_a - {1 \over 2} (
\nabla \cdot\xi) g_{ab}  ,      \equn\put\ckv
$$
leads to the following Jacobian of the change of variables to
$h^{\perp}$, $\xi$, and $\sig$:
$$
J = {\det}^{'{1 \over 2}} (L^{\dag} L),\equn\put\jac
$$
where $L^{\dag}$ is the Hermitian adjoint of $L$ as defined by the inner
product {\innr}. Explicitly,
$$
(L^{\dag} L)^a_{\ b} = -2( \delta^a_{\ b} \sq + {1 \over 2} \nabla^a
\nabla_b
+ R^a_{\ b}). \equn\put\gho
$$
The prime in {\jac} indicates that the zero modes of $L$ must be excluded
from
$J$ and treated separately.
\par
A word about the relationship of this ghost determinant to that obtained by
ordinary Fadeev-Popov gauge fixing is in order. In the standard approach
one fixes a gauge by requiring that the space of deformations spanned
by $h^{\perp}$ satisfy some linear condition of the form,
$$
{\cal F}^b h^{\perp}_{ab} = 0 ,\equn\put\gaug
$$
where $\cal F$ is independent of the general gauge field to be integrated,
in this case the metric. One can easily show [\cBlau] that in gauges
of this form, the Jacobian of the change of variables above becomes
$$
J = {\det}^{-{1 \over 2}} ({\cal F} \circ{\cal F}^{\dag}) {\det} ({\cal F}
 \circ L) \equn\put\jgen
$$
instead of {\jac}. The determinant to the first power is the ordinary
Fadeev-Popov determinant in the gauge {\gaug}, while the other determinant
is a field independent normalization factor which may be removed outside
of the functional integral over metrics.
It is only when the $h^{\perp}$ components of the metric are required
to satisfy the field {\it dependent} condition,
$$
(L^{\dag} h^{\perp})_a = -2 \nabla^b h^{\perp}_{ab} = 0
\equn\put\hper
$$
that the Jacobian is given by {\jac}. This choice of coordinates on
the configuration space of metrics is the natural, orthogonal one
with respect to the inner product {\innr}, but in the more standard
language
of gauge-fixing, it corresponds to a non-linear field dependent
condition on the metric deformations $h^{\perp}$. The signal of this
is that the determinant in {\jac} appears to the one-half power.
\par
Although no particular gauge choice is preferred over any other,
the choice of the coordinates on field space (gauge) satisfying {\hper} is
very useful in the present context. This is because of the transformation
properties of the operators $L$ and $L^{\dag}$ under the substitution
{\confdef}:
$$
L = e^{2 \sig} \bar L e^{-2 \sig}\qquad , \qquad
L^{\dag} = e^{-4 \sig} \bar L^{\dag} e^{2 \sig}\equn\put\Ltrans
$$
Hence $L^{\dag}L$ is the product of two operators each of which
transform covariantly under a local conformal transformation.
Using the heat kernel definition for the determinant and these
transformation properties we find[\cMM]:
$$
\eqalign{
-{1 \over 2}\delta \ln {\det}'  (L^{\dag}L) &= {\trp}{1 \over 2}\delta
\int_{\eps}^{\infty}{ds \over s} e^{-s L^{\dag}L}\cr &=
{\trp}\int_{\eps}^{\infty}ds \Bigl\{ - 2 \delta\sig L^{\dag}L e^{-s
L^{\dag}L} +
2 L^{\dag} \delta\sig L e^{-s L^{\dag}L} -  L^{\dag}L \delta\sig e^{-s
L^{\dag}L} \Bigr\} \cr &={\trp}  \int_{\eps}^{\infty} ds\Bigl\{ - 3
\delta\sig
L^{\dag}L  e^{-s L^{\dag}L} + 2 \delta\sig LL^{\dag} e^{-s LL^{\dag}}
\Bigr\},
\cr &= {\trp} \int_{\eps}^{\infty} ds{d \over ds}
\Bigl\{ 3 \delta\sig e^{-s L^{\dag}L} - 2  \delta\sig e^{-s LL^{\dag}}
\Bigr\} \cr & = - 3 {\trp} \delta\sig e^{-\eps  L^{\dag}L}  +
2 {\trp} \delta\sig e^{-\eps LL^{\dag}}, \cr }  \equn\put\var
$$
where the cyclic property of the trace has been used repeatedly, and
the lower limit of the proper time heat kernel has been regulated
by $\eps$, to be taken to zero in the end. Because of the explicit
appearance of ${\trp}$ over the subspace of nonzero
modes of $L$, the upper limit of the evaluation of
the integral in $s$ vanishes and only the lower limit survives in {\var}.
\par
Here an essential difference from the two dimensional
case manifests itself in the appearance of the {\it tensor} operator
$LL^{\dag}$ whose kernel is {\it infinite} dimensional, being spanned by
all transverse, traceless graviton mode fluctuations. Unlike for
$D= 2$, where the zero modes of $LL^{\dag}$ are countable by
the Riemann-Roch theorem, and their conformal variations may be added
explicitly to {\var}, in $D=4$ these modes cannot be ``counted" without
some action over the transverse, traceless degrees of freedom $h^{\perp}$.
Equivalently, if we exclude these modes by restriction to the
non-zero mode space of $LL^{\dag}$, then the conformal variation of the
ghost
operator $L^{\dag}L$ in {\var} is necessarily non-local, and violates
WZ consistency by itself. The ghost operator cannot yield a coordinate
invariant effective action unless it is combined with the action for
the physical graviton modes. As in the case of determining the Jacobian
{\conjac}, this is another illustration of the fact that the anomaly
coefficients (central charges) cannot be fixed by kinematic considerations
of
functional measures or mode ``counting" alone. Information from the
differential
operator(s) appearing in the Lagrangian, {\it i. e. dynamical} information
about
the theory is required.   \par To see how this works in zeta function
regularization, let the two terms in the last line of {\var} be represented
by
$-3 \zeta (0|L^{\dag}L) \delta \sig$  and $+2\zeta(0|LL^{\dag})\delta \sig$
respectively, for {\it global} conformal variations. The zeta function is
evaluated over the non-zero mode subspaces of each operator.  Since
$LL^{\dag}$
annihalates the trace part of the metric variation, the only tensors in the
general decomposition {\defo} which survive in the second term of {\var}
are
precisely those which are in the range of $L$. Since there is a one-to-one
correspondence between eigenvectors $\xi^{(\lambda)}$ of $L^{\dag}L$ and
eigentensors of $LL^{\dag}$ in the range of $L$ by the relation:
$$
(LL^{\dag})(L\xi^{(\lambda)}) = L(L^{\dag}L)\xi^{(\lambda)} = \lambda
(L\xi^{(\lambda)}), \equn\put\eigen
$$
it follows that their zeta functions are
equal:
$$
\zeta(0|L^{\dag}L) = \zeta (0|LL^{\dag}).\equn\put\zequal
$$
If $L$ and $L^{\dag}$ have a finite number of zero modes these must be
added
explicitly
to the zeta functions to find the full conformal variation of the the two
terms
in {\var}. Then, using {\zequal}, {\var} becomes:
$$
-{1 \over 2}{\delta\over
\delta \sig} \ln \det (L^{\dag}L) = -\zeta (0|L^{\dag}L) - 3
N_0(L^{\dag}L) + 2
N_0(LL^{\dag}), \equn\put\zefor
$$
in four dimensions. In $D$ dimensions the
corresponding formula has ${D \over2} + 1 $ and ${D \over2}$ replacing the
$3$
and $2$ respectively. Since $N_0(LL^{\dag})$ is infinite for $D=4$, this
formula
is meaningless as it stands, and the last term can be defined only by
specifying
some action for the transverse, traceless modes, which leads to its own
zeta
function definition of $N_0(LL^{\dag})$ from the corresponding differential
operator for these modes.  \par The algorithm for the evaluation of the
$a_2$
coefficient of the general (non-minimal) vector operator has been given by
Barvinsky and Vilkovisky [\cBV]. Applying their general formula for the
operator,
$$
M^a_{\ b} = \sq \delta^a_{\ b} - \lambda \nabla^a \nabla_b + P^a_{\
b}, \equn $$
{\it viz.}
$$
\eqalign{
a_2(M) &= {1 \over 48 \pi^2} \Bigl\{-{11 \over 60} G + ({1 \over 8} \ga^2 +
{1 \over 4}\ga - {4 \over 5}) \Ric + ({1\over 16} \ga^2 + {1\over 4} \ga +
{7 \over 20})\R\cr  &\qquad + ({1 \over 4} \ga^2 + \ga) R_{ab}P^{ab} + ({1
\over
8} \ga^2 + {3 \over 4}\ga - {3 \over 2}) P_{ab}P^{ab} \cr & \qquad \qquad +
({1
\over 8} \ga^2 + {1 \over 4}\ga - {1\over 2}) RP + {1 \over 16} \ga^2
P^2\Bigr\},\cr} \equn
$$
to the case of $L^{\dag}L$ in {\gho} with $\ga =
{\lambda \over 1 - \lambda} = - {1 \over 3}$ and $P_{ab} = R_{ab}$, we
obtain
the result: $$ a_2 (L^{\dag}L) = {1 \over (4\pi)^2} \Bigl\{-{11 \over 180}
\Rie
+  {37 \over 135} \Ric + {19 \over 108} \R\Bigr\}, \equn\put\vect
$$
which has a non-vanishing $R^2$ contribution when espressed in the
basis of {\anomf}. This is because $a_2 (L^{\dag}L)$ is {\it not} the
conformal variation of $\ln\det (L^{\dag}L)$ because of the extra
contribution of $N_0(LL^{\dag})$ in {\zefor}. The contribution proportional
to $\R$ in {\vect} should cancel against the regularized
contribution from the transverse, traceless modes $h^{\perp}$
of a classical conformally invariant theory, when expressed in the
basis {\anomf}.

\vfill
\eject

\newcount\eqncount
\sectadv
\set\sect{4}
\beginsection{4. The Transverse Spin-2 Contribution to the Central Charges}

\par
In order to illustrate in a concrete example how the WZ condition works and
the
$R^2$ contribution to the ghost anomaly in {\vect} is cancelled, we
consider
the Weyl-squared action for the graviton degrees of freedom. In
the present context, this action has the advantage of being classically
conformally invariant, so that the WZ condition
may be checked explicitly. Of course, use of this higher derivative
action for the graviton modes leads to perturbative non-unitarity,
about which we have nothing new to add. The restriction to the one-loop
contribution of this action is equivalent to imposing a self-duality
constraint on the graviton degrees of freedom[\cScmh]. Our calculation will
also provide an independent check on those that have appeared in the
literature
on the Weyl-squared theory[\cFT].
\par
The fourth order tensor operator for the linearized Weyl squared
action may be put into the standard form of {\oper} with[\cCD]:
$$
\eqalign{
(\hD_T^{\mu\nu})^{ab}_{\ \  cd} &= (-{2\over 3} R g^{\mu \nu} + 2
R^{\mu\nu})
\h1^{ab}_{\ \  cd}  + {4\over 3}R^{ab}\delta^{(\mu}_c\delta^{\nu)}_d +
4g^{\mu\nu}R^{a\ b}_{\ c\ d} - 4 \delta_c^a \delta^{(\mu}_d R^{\nu)b}\cr
(\hP_T)^{ab}_{\ \ cd} &= ({1\over 3} \R - R_{\mu\nu}^2)\h1^{ab}_{\ \ cd}
- {4 \over 3} R R^{a\ b}_{\ c\ d} - {4\over 3}R^{ab}R_{cd} - {4\over 3}
RR^a_d\delta^b_c \cr &\qquad + 2R^a_cR^b_d + 4 R^a_eR^e_d\delta^b_c +
2 R^e_dR^{a\ b}_{\ c\ e} + 4R^{aebf}R_{cedf}\cr
(\hR_T^{\mu\nu})^{ab}_{\ \ cd} &= R^{a\ \mu\nu}_{\ c}\delta^b_d +
 R^{b\ \mu\nu}_{\ d}\delta^a_c \cr}\equn\put\tensop
$$
\par
In order to use the algorithm {\algo}, we must extend the quartic tensor
operator acting on transverse traceless tensors, spanned by
$h^{\perp}_{ab}$ to
act on all traceless symmetric tensors, {\it i. e.} the union of the spaces
spanned by $h^{\perp}_{ab}$ and $(L\xi)_{ab}$. This we may do by multipling
the
expressions for $\hD_T$ and $\hP_T$ from the left and right by the
traceless
projector $\h1^{(TF)}$ given by:
$$
(\h1^{(TF)})^{ab}_{\ \ cd} = {1 \over
2}(\delta^a_{\ c}\delta^b_{\ d} + \delta^a_{\ d}\delta^b_{\ c} - {1 \over
2}
g^{ab}g_{cd}).\equn\put\trf
$$
Then we must subtract from the $a_2$ of the
resulting tensor operator the four vector modes of the operator ${\cal
M}_V$ we
have added to the transverse tensors. ${\cal M}_V$ is determined by the
property,
$$
{\cal M}_T (L \xi) = L ({\cal M}_V)\xi, \equn\put\ginv
$$
which in the case that ${\cal M}_T$ is given by the standard form {\oper}
with the expressions in {\tensop} is again of the standard form {\oper}
with:
$$
\eqalign{
(\hD_V^{\mu\nu})^a_{\ b} &= (-{2\over 3}Rg^{\mu\nu} + 2 R^{\mu\nu})
\delta^a_b -R^{a(\mu}\delta_b^{\nu)}\cr
(\hP_V)^{a}_{\ b} &= {3 \over 2}R^{ac}R_{bc} - {2\over 3}RR^a_b \cr
(\hR_V^{\mu\nu})^a_{\ b} &= R^a_{\ b\mu\nu}.\cr}\equn\put\qvec
$$
In these formulae, we have systematically neglected all terms involving
derivatives of the Riemann tensor and its contractions, so that $\sq R$
contributions to the anomaly have been dropped. Then, the $a_2$ coefficient
for the extended tensor operator ${\cal M_T}^{(TF)}$ and the vector
operator ${\cal M}_V$ may be computed using the general quartic algorithm
{\algo}. After a straightforward but tedious exercise in index contraction,
we
obtain
$$
a_2({\cal M_T}^{(TF)}) =  {1 \over (4\pi)^2} \Bigl\{{21 \over 10} \Rie
+ {2417 \over 1080}\Ric - {455 \over 432} \R\Bigr\}. \equn\put\tens
$$
and
$$
a_2({\cal M}_V) = {1 \over (4\pi)^2} \Bigl\{-{11\over 90}\Rie - {781\over
360} \Ric + {131 \over 144}\R\Bigr\}. \equn
$$
Subtracting this last quantity and the result for the vector ghost
operator $L^{\dag}L$ of {\vect} above from {\tens} yields finally:
$$
\eqalign{
a_2(F) &= a_2({\cal M}_T^{(TF)}) - a_2({\cal M}_V) - a_2(L^{\dag}L)\cr
&= {1 \over (4\pi)^2} \Bigl\{{199 \over 30}F - {87 \over 20} G\Bigr\},
\cr}\equn\put\resF
$$
thus verifying explicitly the cancellation of $\R$ in the basis {\anomf},
as required by WZ consistency. The coefficients in {\resF} agrees with
the results for the Weyl Lagrangian obtained previously by Fradkin and
Tseytlin[\cFT].
\par
As yet another check of these calculations we may use the $\zeta$ function
technique on the maximally symmetric Euclidean $S_4$ with radius $H^{-1}$.
Decomposing the general vector field $\xi$ on which $L^{\dag}L$ acts into
its
transverse and longitudinal components, we have:
$$
(LL^{\dag})_a^{\ b}
({\xi}^{\perp}_b + \nabla_b \psi) =  2(-\sq - {R \over 4}) {\xi}^{\perp}_a
+ 3
\nabla_a (-\sq - {R \over 3}) \psi \equn
$$
on $S_4$. The transverse vector operator has eigenvalues $(n+4)(n-1)H^2$
with
degeneracy $n(n + {3 \over 2})(n + 3)$. This leads to the zeta function
$$
\zeta_{\perp}(s|L^{\dag}L) = \Bigl( {\mu \over H } \Bigr)^{2s}
\sum_{n=2}^{\infty} {n(n + {3 \over 2})(n + 3) \over (n+4)^s (n-1)^s},\equn
$$
where the ten zero modes (Killing vectors of $S_4$) at $n=1$ are excluded
from the sum. Analytically continuing this zeta function to $s=0$ by
the technique already described above in connection with {\zedel} yields:
$$
\zeta_{\perp}(0|L^{\dag}L) = \zeta_R (-3) + {15 \over 2} \zeta_R (-2) +
{33 \over 2} \zeta_R (-1) + 10 \zeta_R (0) = - {191 \over 30}
\equn\put\zetr
$$
in agreement with ref. [\cCD]. The scalar operator $-\sq - 4 H^2$ has
the same eigenvalue spectrum as the transverse vector
operator on $S_4$ with the scalar degeneracy, ${1 \over 3}
(n + 1) (n + {3 \over 2})(n + 2)$. The five zero modes (conformal
Killing vectors of $S_4$) at $n=1$ are excluded. Thus the total number of
zero modes of $L^{\dag}L$ on the sphere is:
$$
N_0(L^{\dag}L) = 15
\equn\put\zerom
$$
The constant mode with negative eigenvalue at $n=0$ is not in the spectrum
of
$L^{\dag}L$, since the gradient of a constant vanishes identically.
The scalar's $\zeta$ function,
$$
\zeta_S(s|L^{\dag}L) = {1 \over 3} \Bigl( {\mu \over H } \Bigr)^{2s}
\sum_{n=2}^{\infty}{(n + 1)(n + {3 \over 2})(n + 2) \over
(n+4)^s (n-1)^s} , \equn
$$
evaluated by the same technique at $s=0$ yields:
$$
\zeta_S(0|L^{\dag}L) = {1 \over 3}\zeta_R (-3) + {5 \over 2} \zeta_R (-2)
 + {37 \over 6} \zeta_R (-1) + 5 \zeta_R (0) = - {271 \over 90}.
\equn\put\zets
$$
\par
The quartic operator ${\cal M}_T$ factorizes on $S_4$ into the product,
$(-\sq + 2H^2)(-\sq + 4H^2)$ with eigenvalues $H^4 n(n+1)(n+2)(n+3)$,
and tensor degeneracy ${5 \over 6}(n-1)(n+4)(2n+3)$ giving:
$$
\zeta_{\perp}(s|{\cal M}_T) = {5 \over 6} \Bigl( {\mu \over H } \Bigr)^{4s}
\sum_{n=2}^{\infty}{(n - 1)(n + 4)(2n + 3) \over
n^s(n+1)^s (n+2)^s (n+3)^s},\equn
$$
with the result,
$$
\zeta_{\perp}(0|{\cal M}_T) = {164 \over 9},\equn
$$
which may be obtained as well by taking the sum of the zeta functions
at $s=0$ of the two factorized second order operators. Using this last
value of $164/9$ as the regulated $2N_0(LL^{\dag})$ appearing in
eq. {\zefor}, we obtain the total scaling behavior of the Weyl
theory on the maximally symmetric space $S_4$ (where $F=0$):
$$
-{1 \over 2}\delta \ln \det (L^{\dag}L) =  {191 \over 30} + {271 \over 90}
-3\cdot 15 + {164 \over 9} = -{87 \over 5},\equn
$$
which gives:
$$
b'_G = -{1 \over (4\pi)^2}{87 \over 20},\equn
$$
in perfect agreement with {\resF}. In units of $Q^2$ this gives
${87 \over 10} = 8.7$ for the spin-2 contribution to the central
charge appearing in the anomalous scaling relation {\alp} of
the effective $\sig$ theory.
\par
We have considered the Weyl-squared Lagrangian primarily as a good
illustrative
example of the WZ condition. However, it is far from clear that this is the
appropriate action to use for computing the contributions of the
transverse,
traceless graviton degrees of freedom to the central charges of the
effective
$\sig$ theory in the infrared, which is the regime we may hope to apply
this
theory to four dimensional physics. If we attempt to use the Einstein
theory for
this purpose we run into several well-known difficulties. The first problem
is
that loop calculations in quantum gravity, as in non-Abelian gauge theory,
make
sense only if the the  background field equations are satisfied. However,
the
field equations  of the Einstein theory obscure the WZ condition, since
they
imply  that any $R^2$ term in the anomaly is indistinguishable from $\Ric$,
 on
shell. One possible way out of this difficulty is to use the conformal
off-mass
shell extension of Fradkin and Vilkovisky[\cFV]. More serious, of course,
is the
fact that the Einstein theory is non-renormalizable. Hence it is not clear
that
we can trust any one-loop calculation of $a_2$ coefficients in the Einstein
theory for  the $Q^2$ and anomalous scaling behavior we find in the
effective
$\sig$ theory, (which {\it is} renormalizable). \par If we simply ignore
these
difficulties, the one loop calculation in the Einstein theory reduces to
computing the $a_2$ coefficient of the Lichnerowicz laplacian,
$$
({\cal M}_T)^{ab}_{\ \ cd} = \sq {\hat 1}^{ab}_{\ \ cd} +
2 R^{a\ b}_{\ c\ d}\equn\put\Lic
$$
over transverse, traceless tensors. In order to use the standard
algorithm for general second order operators, we again extend ${\cal M}_T$
to act on all traceless tensors, by
multiplying it on both sides by the traceless projector, {\trf}.
Then we subtract from the $a_2$ of this operator the four
vector modes of the operator ${\cal M}_V$ we have added to the
transverse tensors. In this case ${\cal M}_V$, determined from {\ginv} is
given by:
$$
({\cal M}_V)^a_{\ b} = \sq \delta^a_{\ b} + R^a_{\ b},\equn
$$
Then the standard algorithm gives:
$$
\eqalign{
a_2({\cal M}_T^{(TF)}) &= {21\over 20}(\Rie - \Ric) \cr
-a_2({\cal M}_V) &= {11\over 180}\Rie - {43 \over 90}\Ric -
{2 \over 9}\R. \cr} \equn
$$
for the Einstein theory at one-loop order.
When the ghost contribution of {\vect} is subtracted from the sum of
these, and the vacuum Einstein equations, $R_{ab} = R g_{ab} /4$ are used,
we find:
$$\eqalign{
b_G &= \ \ {1\over (4 \pi)^2}\ {\ 611\over 120}\cr
b'_G &= -{1\over (4 \pi)^2}\  {1411\over 360} \cr}\equn\put\resE
$$
These results agree with the standard results for the Einstein theory
with the conformal part removed[\cCD]. This gives a contribution to $Q^2$
of $1411/180 \approx 7.9$, which curiously does not differ much from
that obtained in the Weyl-squared theory, {\it viz.} $8.7$
\par
In either case, it is noteworthy that the total $b$ and $-b'$ coefficients
are positive, and dominated by the ghost + graviton contributions,
which add with the same sign as the matter contributions.
This is different from the two dimensional result that
the matter and ghosts contribute to the central charge
with opposite sign. Inspection of the quartic action
{\actan} reveals that the ghosts behave the same way
in two and four dimensions. It is the {\it matter}
which behaves differently. Namely, in two dimensions
the matter contributes to a negative kinetic term for the
Liouville theory, but in four dimensions the matter contribution
to the fourth order action is positive in Euclidean
signature. Although a general proof is lacking, it seems
that $Q^2$ is always positive. As already remarked above,
both the Einstein and Weyl squared calculations are not
entirely free of problems, so that we regard the issue of the correct
contribution of the spin-2 modes to the value of $Q^2$
in the infrared as still open, and deserving of further study.

\vskip1.5cm
\centerline{\it Acknowledgements}
\vskip.5cm

This work was supported in part by the NATO grant CRG900636.
I.A. and P.O.M, and E.M. thank respectively Los Alamos National Laboratory
and
Ecole Polytechnique for hospitality during the completion of this work.

\baselineskip = 15pt

\parindent=-10 pt

\vfill
\eject
\centerline{\bf REFERENCES }
\vskip 1cm

\parskip=-3 pt

\item{\cPoly}. A. M. Polyakov, {\it Phys. Lett.} {\bf B103} (1981)
 207.\hfill\break

\item{\cDuf}. M. J. Duff, {\it Nucl. Phys.} {\bf B125} (1977) 334;
\hfill\break\indent
N. D. Birrell and P. C. W. Davies, {\it
``Quantum Fields in Curved Space,"} (Cambridge \hfill\break Univ. Press,
 Cambridge, 1982)
and references therein.\hfill\break

\item{\cWZ}. J. Wess and B. Zumino, {\it Phys. Lett.} {\bf B37} (1971) 95;
\hfill\break\indent
L. Bonora, P. Cotta-Rasmusino, and C. Reina, {\it Phys. Lett.}
{\bf B126} (1983)
305.\hfill\break

\item{\cAntMot}. I. Antoniadis and E. Mottola, {\it Phys. Rev.} {\bf D}
(1992).
\hfill\break

\item{\cOdR}. R. J. Riegert, {\it Phys. Lett.} {\bf B134} (1984) 56;
\hfill\break\indent
E. S. Fradkin and A. A. Tseytlin, {\it Phys. Lett.} {\bf B134} (1984) 187;
\hfill\break\indent
E. T. Tomboulis, {\it Nucl. Phys.} {\bf B329} (1990) 410;
\hfill\break\indent
S. Odintsov and I. Shapiro, {\it Class. Quant. Grav.} {\bf 3} (1991).
\hfill\break

\item{\cMM}. P. O. Mazur and E. Mottola, {\it Nucl. Phys.} {\bf B341}
(1990)
 187;\hfill\break\indent
P. O. Mazur, {\it Phys. Lett.} {\bf B262} (1991) 405.\hfill\break

\item{\cAnt}. R. C. Myers, {\it Phys. Lett.} {\bf B199} (1987)
 371;\hfill\break\indent
I. Antoniadis, C. Bachas, J. Ellis and D. V. Nanopoulos, {\it Nucl. Phys.}
{\bf
 B328} (1989) 117.\hfill\break

\item{\cBV}. A. O. Barvinsky and G. A. Vilkovisky, {\it Phys. Rep.} {\bf
119}
 No.1 (1985) 1. \hfill\break

\item{\cDeW}. B. S. DeWitt, {\it ``Relativity, Groups, and Topology,"} B.
S. DeWitt and C. DeWitt, eds. (Gordon and Breach, New York, 1964); {\it
``General Relativity: An Einstein Centenary Survey,"} S. W. Hawking and W.
Israel, eds. (Cambridge Univ. Press, Cambridge, 1979);\hfill\break\indent

\item{\cBlau}.  S. K. Blau, Z. Bern, and E. Mottola, {\it Phys. Rev.}
{\bf D43} (1991). 1212.\hfill\break

\item{\cFV}. E. S. Fradkin and G. A. Vilkovisky, {\it Phys. Lett.}
{\bf B73} (1978) 209. \hfill\break

\item{\cCD}. S. M. Christensen and M. J. Duff, {\it Nucl. Phys.} {\bf B170
[FS1]} (1980) 480; \hfill\break\indent
E. Fradkin and A. A. Tseytlin, {\it Nucl.Phys.} {\bf B234} (1984) 509.
\hfill\break

\item{\cScmh}. C. Schmidhuber, CalTech report CALT-68-1745
(1992).\hfill\break

\item{\cBC}. N. H. Barth and S. M. Christensen, {\it Phys. Rev.} {\bf D28}
(1983) 1876. \hfill\break

\item{\cFT}. E. Fradkin and Tseytlin {\it Phys. Lett.} {\bf B104} (1981)
377; {\it Nucl. Phys.} {\bf B201} (1982) 469; {\it Phys. Lett.} {\bf B110}
(1982) 117; \hfill\break\indent
S. Odintsov, to appear in {\it Zeit. Phys.} (1992). \hfill\break\indent

\end